\documentclass[aps,pra,preprint,superscriptaddress,longbibliography]{revtex4-1}
\usepackage{amsmath}
\usepackage{amsthm}
\usepackage{amsfonts}
\usepackage{amssymb}
\usepackage{xcolor,graphicx}
\usepackage{tikz}
\usepackage{color}
\usepackage{hyperref}
\hypersetup{
	colorlinks=true,
	linkcolor=blue,
	filecolor=magenta,      
	urlcolor=blue,
}
\usepackage{longtable}
\usepackage{array}
\usepackage{dsfont}

\begin{document}
	
%--------------------------------------------------------------------------------
	\title{Engineering $\mathrm{SU}(1,1) \otimes \mathrm{SU}(1,1)$ vibrational states}
	
	\author{C. Huerta Alderete}
	\email[e-mail: ]{aldehuer@inaoep.mx}
	\affiliation{Instituto Nacional de Astrof\'{\i}sica, \'Optica y Electr\'onica, Calle Luis Enrique Erro No. 1, Sta. Ma. Tonantzintla, Pue. CP 72840, M\'exico}

	\author{M. P. Morales Rodr\'iguez}
	\affiliation{Programa Delf\'in, Verano de la Investigaci\'on Cient\'ifica -- Instituto Nacional de Astrof\'{\i}sica, \'Optica y Electr\'onica, Calle Luis Enrique Erro No. 1, Sta. Ma. Tonantzintla, Pue. CP 72840, M\'exico}

	\author{B. M. Rodr\'iguez-Lara}
	\email[e-mail: ]{bmlara@tec.mx}
	\affiliation{Instituto Nacional de Astrof\'{\i}sica, \'Optica y Electr\'onica, Calle Luis Enrique Erro No. 1, Sta. Ma. Tonantzintla, Pue. CP 72840, M\'exico}
	\affiliation{Tecnologico de Monterrey, Escuela de Ingenier\'ia y Ciencias, Ave. Eugenio Garza Sada 2501, Monterrey, N.L., M\'exico, 64849.}

 	\date{\today}
	
	\begin{abstract}
	We propose an ideal scheme for preparing vibrational $\mathrm{SU(1,1)} \otimes \mathrm{SU(1,1)}$ states in a two-dimensional ion trap using red and blue second sideband resolved driving of two orthogonal vibrational modes.
	Symmetric and asymmetric driving provide two regimes to realize quantum state engineering of the vibrational modes. 
	In one regime, we show that time evolution synthesizes so-called $\mathrm{SU}(1,1)$ Perelomov coherent states, that is separable squeezed states and their superposition too.
	The other regime allows engineering of lossless 50/50 $\mathrm{SU}(2)$ beam splitter states that are entangled states.
	These ideal dynamics are reversible, thus, the non-classical and entangled states produced by our schemes might be used as resources for interferometry.
	\end{abstract}
	
	%\pacs{05.45.Mt, 42.50.Ct, 42.50.Mn, 73.43.Nq}
	
	\maketitle
%--------------------------------------------------------------------------------
%%%%%%%%%%%%%%%%%%%%%%%%%%  body  %%%%%%%%%%%%%%%%%%%%%%%%%%
\section{Introduction}
Quantum state engineering studies the preparation, manipulation, and characterization of arbitrary quantum states. 
Technological advances allow the coherent control of dynamics in an increasing collection of physical systems; e.g. trapped ions, superconducting circuits, quantum gases, mechanical oscillators.
In particular, trapped ions show high addressability, long coherence times and high fidelity readout necessary for quantum state preparation and manipulation within its own experimental issues \cite{Wineland1998}.
Vibrational state characterization is available in this platform through tomographic reconstruction, experimentally demonstrated for Wigner \cite{Leibfried1996} and Hussimi \cite{Lv2017} quasi-probability distributions.

Trapped ions have proved a fertile ground for fundamental research and quantum technologies development \cite{Kielpinski2002,Blatt2012}. 
Single-mode vibrational number, coherent, and squeezed states have been engineered experimentally \cite{Meekhof1996} and theoretical proposals for the synthesis of arbitrary one-\cite{Law1996} and two-dimensional vibrational states \cite{Drobny1998} has been produced.
In particular, trapped ions might act as vibrational beam splitters \cite{Gou1996} producing states identical to the photon states on the output ports of a lossless interferometer with number-state inputs \cite{Campos1989}.
Vibrational interferometry can be used to either explore fundamental quantum mechanics, e.g. quantum decoherence \cite{Poyatos1996,Zeng1998}, or produce new quantum technologies, e.g. vibrational thermometers \cite{Johnson2015} or quantum gyroscopes \cite{Campbell2017}.

Squeezing and entanglement improve phase sensitivity in interferometry in a manner proportional to the inverse of the excitation number of quanta entering an interferometer \cite{Yurke1986,Boehmer1995,Carranza2012p2581}.
Here, we are interested in the quantum state engineering of orthogonal vibrational modes that show squeezing and entanglement with an underlying $\mathrm{SU}(1,1) \otimes \mathrm{SU}(1,1)$ symmetry.
We will use blue and red resolved second sideband driving \cite{DeMatosFilho1994,DeMatosFilho1996} to this end. 
In the following, we will present an effective Hamiltonian describing our proposal in the Lamb-Dicke regime.
Then, we will show that the asymmetric coupling model produces the superposition of separable squeezed vibrational mode states where the inner state is intrinsically entangled to the vibrational modes.
Afterward, we will show that red driving with symmetric coupling is able to produce lossless 50/50 $\mathrm{SU}(2)$ beam splitter vibrational states that are factorized from the internal state of the ion.
The ideal dynamics producing these states are reversible and seem to suggest their use as interferometers to characterize different aspects of real-world experiments.

\section{Dual-mode, second-sideband driving model}

We suggest driving two normal phonon modes of the center of mass motion of a trapped ion \cite{Messina2003p1,Zhu2006}, such that the effective Hamiltonian,
\begin{eqnarray}
\hat{H}_{ion} &=& \frac{\omega_{0}}{2}\hat{\sigma}_{3} + \sum_{j=1}^{2} \nu_{j} \hat{a}^{\dagger}_{j} \hat{a}_{j} + \Omega_{j} \cos \left[\eta_{j}\left(\hat{a}^{\dagger}_{j} + \hat{a}_{j}\right) - \omega_{j} t + \phi_{j}\right]\hat{\sigma}_{j},
\end{eqnarray}
describes the interaction of the $j$-th vibrational mode, with frequency $\nu_{j}$ and represented by the annihilation (creation) operator $\hat{a}_{j}$ ($\hat{a}_{j}^{\dagger}$), with two internal levels of the trapped ion, with energy gap $\omega_{0}$ and represented by Pauli matrices $\hat{\sigma}_{j}$, through a set of external driving fields of frequency $\omega_{j}$, Lamb-Dicke parameter $\eta_{j}$, phase $\phi_{j}$, and Rabi coupling strength $\Omega_{j}$.
Moving into the reference frame defined by the uncoupled Hamiltonian, $\hat{H}_{0} = \omega_{0} \hat{\sigma}_{3} / 2 + \nu_{1} \hat{a}_{1}^{\dagger} \hat{a}_{1} + \nu_{2} \hat{a}_{2}^{\dagger} \hat{a}_{2}$, driving the $k$-th vibrational sideband, $\omega_{j} = \omega_{0} - k \nu_{j}$ where $k>0$ and $k<0$ define the so-called red and blue driving, in the Lamb-Dicke regime, $\eta_{j} \sqrt{ \langle \hat{a}_{j}^{\dagger} \hat{a}_{j} \rangle } \ll 1$, and under an optical and mechanical rotating wave approximation, we can approximate \cite{Vogel1995} for red,
\begin{eqnarray}
\hat{H}_{R} \approx \sum_{j=1}^{2} \frac{g_{j,k}}{2} \left[e^{i \phi_{j,k}} \hat{a}_{j}^{k} \hat{\sigma}_{+} + e^{-i \phi_{j,k}} \hat{a}_{j}^{\dagger k} \hat{\sigma}_{-}\right],
\end{eqnarray}
and blue sideband driving,
\begin{eqnarray}
\hat{H}_{B} \approx \sum_{j=1}^{2} \frac{g_{j,k}}{2} \left[e^{i \phi_{j,k}} \hat{a}_{j}^{\dagger \vert k \vert} \hat{\sigma}_{+} + e^{-i \phi_{j,k}} \hat{a}_{j}^{\vert k \vert } \hat{\sigma}_{-}\right],
\end{eqnarray}
with effective couplings and phases, $g_{j,k} \approx \Omega_{j} \eta_{j}^{\vert k \vert } e^{- \vert \eta_{j} \vert^{2}/2} /\vert k \vert !$ and $e^{i \phi_{j,k}} =  (-i)^{j-1 + \vert k \vert } e^{i\phi_{j}}$, in that order.
For reasons that will become obvious, we choose the second sideband driving \cite{DeMatosFilho1996}, $k=\pm 2$, and draw from the idea of simultaneous blue and red driving in the simulation of the quantum Rabi model \cite{Pedernales2015,Lv2018} to reach the model Hamiltonian at the core of our proposal,
\begin{eqnarray}
\hat{H} &=& \sum_{j=1}^{2} \frac{g_{j,2}}{2} \left[e^{i \phi_{j,2}} \hat{a}_{j}^{2} \hat{\sigma}_{+} + e^{-i \phi_{j,2}} \hat{a}_{j}^{\dagger 2} \hat{\sigma}_{-}\right] +  \frac{g_{j,-2}}{2} \left[e^{i \phi_{j,-2}} \hat{a}_{j}^{\dagger 2} \hat{\sigma}_{+} + e^{-i \phi_{j,-2}} \hat{a}_{j}^{ 2 } \hat{\sigma}_{-}\right].
\end{eqnarray}
This model is the two-phonon interaction analog of the so-called cross-cavity quantum Rabi model \cite{HuertaAlderete2016p414001} that has been used to propose the quantum simulation of para-oscillators in trapped ions \cite{HuertaAlderete2017p013820,HuertaAlderete2018p11672}.
Single-mode, two-phonon interactions in trapped ions have been recently proposed to simulate interaction-induced spectral collapse\cite{Felicetti2015p033817}.
It may be possible to explore alternative schemes in the trapped-ion platform; for example, continuous dynamical decoupling schemes that have been proposed as alternatives to produce robust realizations of the two-phonon interaction in the ultrastrong regime\cite{Puebla2017p063844} or dynamics far away from the Lamb-Dicke regime\cite{Cheng2018p023624}.
It might be even possible to explore realizations with superconducting qubits, where a two-photon quantum Rabi model has been proposed\cite{Felicetti2018p013851}, or single trapped cold atoms, where a proposal to realize the quantum Rabi model has arisen\cite{Schneeweiss2018p021801}. 

%%%%%%%%%
\section{Asymmetric squeezing}

\begin{figure}[htbp]
	\centering
	\includegraphics[scale=1]{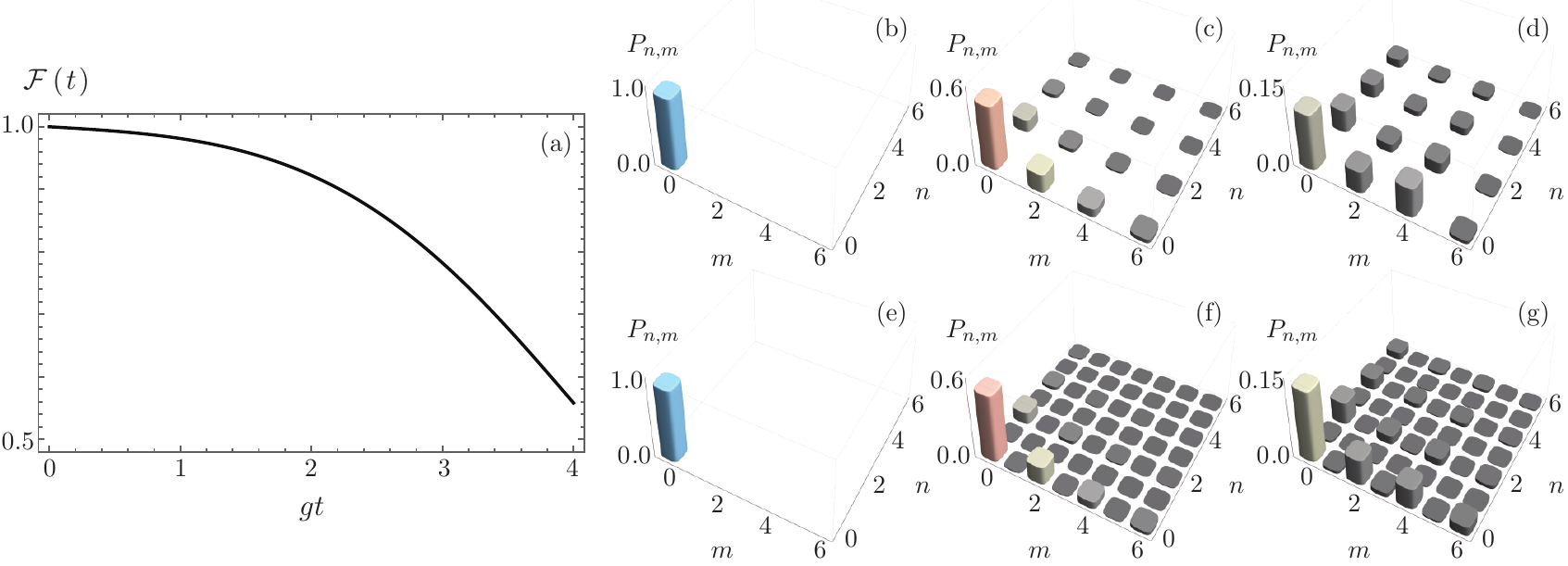}
	\caption{(a) Fidelity between ideal evolution and evolution for asymmetric $\mathrm{SU}(1,1)$ coherent states with $g_{2} = 0.75~ g_{1}$ and $\gamma = 0.1 ~g_1$. The joint phonon number probability distribution for (b)-(d) ideal and (e)-(g) lossy evolution at (b),(e) $g_{1}t=0$, (c),(f) $g_{1}t=1$, and (d),(g) $g_{1}t=2$ for an initial state $\vert \psi(0) \rangle = \vert 0,0\rangle \left( \vert g \rangle + \vert e \rangle \right)/\sqrt{2}$. }
	\label{fig:Fig1}
\end{figure}

Let us simplify and find uses for our general Hamiltonian. 
First, we propose to work with a system where the amplitudes and relative phases for blue and red driving in each mode are chosen to provide similar coupling parameters, $g_{j,k} \equiv g_{j}$, and phases, $\phi_{j,k} \equiv \pi/2 $.
Under this assumption, the Hamiltonian describing the system,
\begin{eqnarray}
\hat{H}_{1} = i \left[ \frac{g_{1}}{2} \left( \hat{a}^{\dagger 2}_{1} - \hat{a}^{2}_{1}\right) + \frac{g_{2}}{2} \left( \hat{a}^{\dagger 2}_{2} - \hat{a}^{2}_{2}\right) \right] \hat{\sigma}_{1},
\end{eqnarray}
yields an evolution operator,
\begin{eqnarray}
\hat{U}_{1}(t) = \hat{S}_{1}(g_{1}t \hat{\sigma}_{1})\hat{S}_{2}(g_{2}t \hat{\sigma}_{1}),
\end{eqnarray}
that is the product of two standard $\mathrm{SU}(1,1)$ squeezing operators, $\hat{S}_{j}(\alpha) = \mathrm{exp}\left[ \alpha \left( \hat{a}_{j}^{\dagger2} - \hat{a}_{j}^{2} \right)/2 \right]$ controlled by the internal state of the trapped ion.
For example, choosing an initial state with arbitrary phonon fields and the internal state an eigenstate of the $\hat{\sigma}_{1}$ operator, $\vert \psi(0) \rangle = \vert\xi_{1}, \xi_{2}, x_{\pm} \rangle$, the ideal evolution,
\begin{eqnarray}
\vert \psi(t) \rangle = \hat{S}_{1}(\pm g_{1}t) \hat{S}_{2}(\pm g_{2}t) \vert\xi_{1}, \xi_{2}, x_{\pm} \rangle,
\end{eqnarray}
provides a separable state with different squeezing in each mode.
These are the so-called two-mode $\mathrm{SU}(1,1)$ Perelomov coherent states, whose properties have been discussed in detail by Gerry and Benmoussa\cite{Gerry2000}.
Figure \ref{fig:Fig1}(a) shows the fidelity defined as the trace distance,
\begin{eqnarray}
\mathcal{F}(t) = \mathrm{Tr} \left[ \hat{\rho}_{\psi}(t) \hat{\rho}_{\gamma}(t) \right],
\end{eqnarray}
between the density operator describing the ideal evolution, $\hat{\rho}_{\psi}(t) = \vert \psi(t) \rangle \langle \psi(t) \vert$, and that for evolution under atomic decay, $\hat{\rho}_{\gamma}(t)$ such that $\dot{\rho} = - i \left[ \hat{H}, \hat{\rho} \right] + \gamma_{q} \left( \hat{\sigma}_{-} \hat{\rho} \hat{\sigma}_{+} - \left\{ \hat{\sigma}_{+} \hat{\sigma}_{-}, \hat{\rho} \right\}/2 \right) + \sum_{j} \gamma_{j} \left( \hat{a}_{j} \hat{\rho} \hat{a}_{j}^{\dagger} - \left\{ \hat{a}_{j}^{\dagger} \hat{a}_{j}, \hat{\rho} \right\}/2 \right)$ where $\gamma_{q}$ and $\gamma_{j}$ are the effective atomic and vibrational decay rates.
The parameters in the simulations are $g_{2} = 0.75 ~g_{1}$ and, for the sake of the example, homogeneous decays $\gamma_{q} = \gamma_{j} = \gamma = 0.1~g_{1}$.
While we select these parameter values for the sake of showing the dynamics in a lossy system, they are not far from those of trapped Barium\cite{Dietrich2010p052328} or Cadmium\cite{Deslauriers2004p043408} where Rabi frequencies are of the order of 15--100 kHz and the coherence time of the qubits are in the 80--120 $\mu$s range. 
Note that 
Figures \ref{fig:Fig1}(b) to Fig. \ref{fig:Fig1}(d) show the joint phonon number probability,
\begin{eqnarray}
P_{n,m}(t) = \mathrm{Tr}\left[ \vert m, n \rangle\langle m, n\vert \hat{\rho} \right],
\end{eqnarray}
for ideal and Fig. \ref{fig:Fig1}(e) to Fig. \ref{fig:Fig1}(g) for lossy evolution at different times. 
We colored the probability bars and show only a small section of the plot to make comparison simpler . 

\begin{figure}[htbp]
	\centering
	\includegraphics[scale=1]{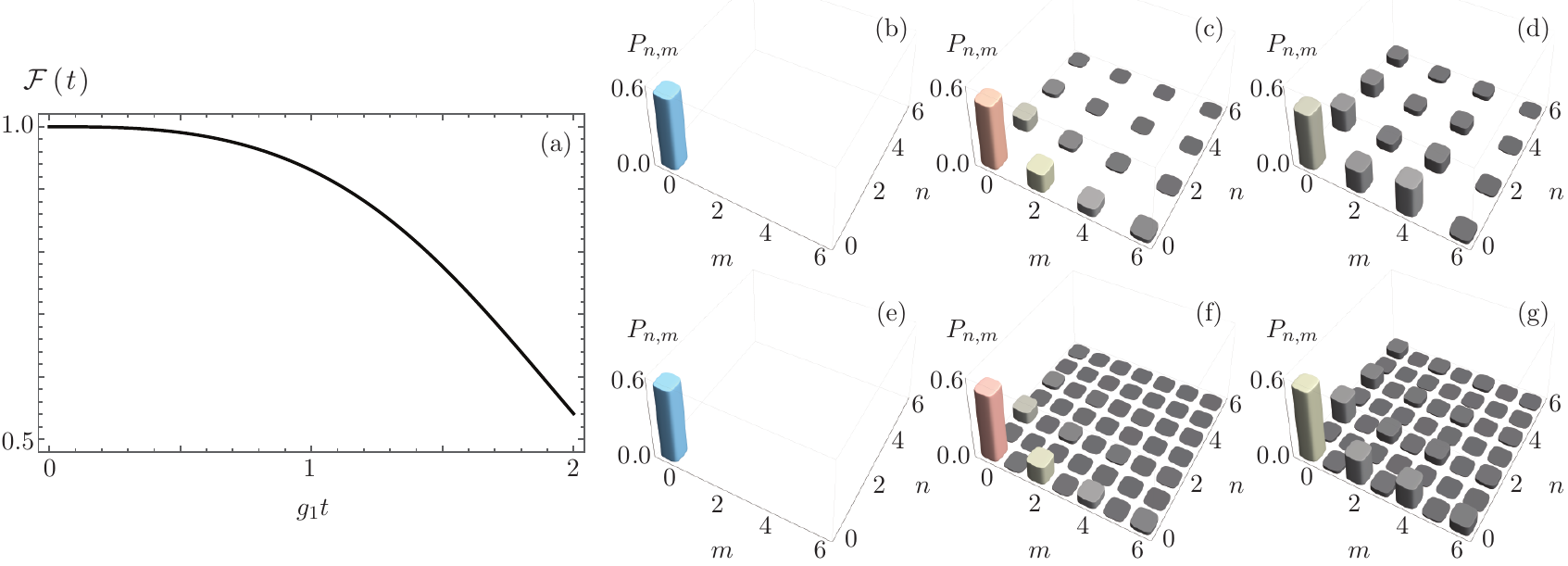}
	\caption{(a) Fidelity between ideal evolution and evolution for asymmetric $\mathrm{SU}(1,1)$ coherent states with $g_{2} = 0.75~g_{1}$ and $\gamma = 0.1 ~g_{1}$. The joint phonon number probability distribution for (b)-(d) ideal and (e)-(g) lossy evolution at (b),(e) $g_{1}t=0$, (c),(f) $g_{1}t=1$, and (d),(g) $g_{1}t=2$ for an initial state $\vert \psi(0) \rangle = \vert 0,0,g \rangle$.}
	\label{fig:Fig2}
\end{figure}

Obviously, choosing an arbitrary internal initial state, $\vert \psi(0) \rangle = \vert\xi_{1}, \xi_{2} \rangle \left( \alpha \vert x_{+} \rangle + \beta \vert x_{-}  \rangle \right)$, provides a superposition of the form,
\begin{eqnarray}
\vert \psi(t) \rangle = \alpha \hat{S}_{1}(g_{1}t) \hat{S}_{2}(g_{2}t) \vert\xi_{1}, \xi_{2}, x_{+} \rangle - \beta \hat{S}_{1}^{\dagger}(g_{1}t) \hat{S}_{2}^{\dagger}(g_{2}t) \vert\xi_{1}, \xi_{2}, x_{-} \rangle, 
\end{eqnarray}
where the state of the whole system is entangled thanks to the internal state of the ion.
Figure \ref{fig:Fig2}(a) shows the fidelity between the ideal and lossy evolution.
Figures \ref{fig:Fig2}(b) to Fig. \ref{fig:Fig2}(d) show the joint phonon number probability, for ideal and Fig. \ref{fig:Fig2}(e) to Fig. \ref{fig:Fig2}(g) for lossy evolution. 
Parameters are identical to those in Fig. \ref{fig:Fig1}.

The fact that a $\pi$-phase change in the driving laser phases ideally reverts the effective model dynamics might suggest its use as a squeezed state interferometer. 
In a real-world experiment, the system will not return to the original initial state due to all the real-world experimental subtleties.
For example, these states might help in the characterization of the environment effect on the trapped center of mass motion and internal states of the ion might mention the obvious.

\section{Symmetric squeezing}

Now, let us simplify our model one step more and assume a system where the coupling parameters, $g_{j,k} \equiv g$, and the phases are chosen to be equal.
The dynamics are given by a simpler Hamiltonian that produces symmetric squeezing if we follow the procedure introduced before. 
We can take a second step and suppress blue sideband driving, then the dynamics are described by a Hamiltonian,
\begin{eqnarray}
\hat{H}_{2} = \frac{g}{2} \left[ \left( \hat{a}^{\dagger 2}_{1} + \hat{a}^{\dagger 2}_{2} \right) \hat{\sigma}_{-} + \left( \hat{a}^{2}_{1} + \hat{a}^{2}_{2} \right) \hat{\sigma}_{+} \right].
\end{eqnarray}
Again, it is straightforward to construct an evolution operator,
\begin{eqnarray}
\hat{U}_{2}(t) = \left(
\begin{array}{cc}
\cos gt \sqrt{\hat{K}_{-} \hat{K}_{+}} & - i \sin gt \sqrt{\hat{K}_{-} \hat{K}_{+}} ~  \frac{1}{\sqrt{\hat{K}_{-} \hat{K}_{+}}} \hat{K}_{-}\\
- i \hat{K}_{+} \frac{1}{\sqrt{\hat{K}_{-} \hat{K}_{+}}} \sin gt \sqrt{\hat{K}_{-} \hat{K}_{+}}  & \cos gt \sqrt{\hat{K}_{+} \hat{K}_{-}}
\end{array}\right),
\end{eqnarray}
using a representation of $\mathrm{SU}(1,1)$,
\begin{eqnarray}
\left[ \hat{K}_{+}, \hat{K}_{-} \right] = - 2 \hat{K}_{3}, \qquad \left[ \hat{K}_{3}, \hat{K}_{\pm} \right] = \pm \hat{K}_{\pm},
\end{eqnarray}
provided by the two-mode operators \cite{Gerry2000},
\begin{eqnarray}\label{eq:algebra2}
\hat{K}_{3} = \frac{1}{2} \left(\sum_{j=1}^{2}\hat{a}_{j}^{\dagger} \hat{a}_{j} + 1\right), \quad 
\hat{K}_{+} =   \frac{1}{2} \left(\hat{a}_{1}^{\dagger 2} +  \hat{a}_{2}^{\dagger 2} \right), \quad
\hat{K}_{-} =  \frac{1}{2} \left(\hat{a}_{1}^{2} +  \hat{a}_{2}^{2} \right).
\end{eqnarray}

\begin{figure}[htbp]
	\centering
	\includegraphics[scale=1]{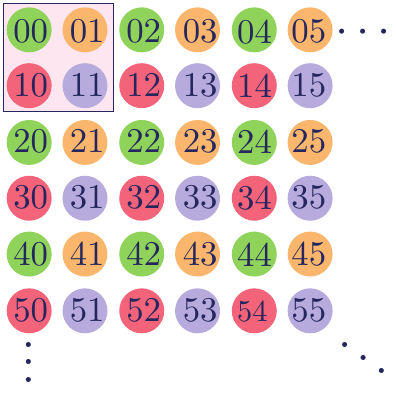}
	\caption{Pictorial representation of the phonon subspaces; the vacuum state for each subspace are located within the marked square. Each color represents a given phonon subspace.}
	\label{fig:Fig3}
\end{figure}

We will use a Hilbert space partition defined by the raising operator,
\begin{eqnarray}
\vert k; m \rangle = \sqrt{ \frac{(2k-1)!}{m! (2k+m-1)!} }~ \hat{K}_{+}^{m} \vert k; 0 \rangle,
\end{eqnarray}
acting on four states that we will call vacuum states related to a Bargmann index $k$.
This produces four phonon subspaces labelled by a Bargmann index and a vacuum state: $k=1/2$ and $\vert 1/2; 0 \rangle = \vert 0,0 \rangle$, $k=1$ and $\vert 1; 0 \rangle_{01} = \vert 0,1 \rangle$, $k=1$ and $\vert 1; 0 \rangle_{10} = \vert 1,0 \rangle$, $k=3/2$ and $\vert 1, 0 \rangle = \vert 1,1 \rangle$.
Figure \ref{fig:Fig3} shows a pictorial representation of these phonon subspaces.
The whole Hilbert state for the quantized center of mass motion is covered once with the four orthogonal subspaces defined by the bases, 
\begin{eqnarray}
\vert 1/2; m \rangle  &=& \frac{1}{2^{m} m!} \sum_{l=0}^{m} \binom{m}{l} \sqrt{(2m - 2l)!(2l)!} ~\vert 2m - 2l, ~2l \rangle , \nonumber \\
\vert 1; m \rangle_{01} &=& \frac{1}{2^{m }} \sum_{l=0}^{m } \binom{m}{l} \sqrt{\frac{(2m -2l+1)!(2l)!}{m! (m + 1)!}} ~\vert 2l,~ 2m - 2l + 1 \rangle, \nonumber \\
\vert 1; m \rangle_{10} &=& \frac{1}{2^{m}} \sum_{l=0}^{m} \binom{m}{l} \sqrt{\frac{(2m-2l+1)!(2l)!}{m! (m +1)!}} ~\vert 2m -2l +1, ~2l\rangle, \nonumber \\
\vert 3/2; m\rangle &=& \frac{1}{2^{m}} \sum_{l=0}^{m} \binom{m}{l} \sqrt{\frac{2 (2m - 2l + 1)!(2l + 1)!}{m! (m + 2)!}} ~\vert 2m-2l +1, ~2l+1 \rangle, 
\end{eqnarray}
that are eigenstates of the $\hat{K}_{3}$ operator, $\hat{K}_{3} \vert k; m \rangle = (m+k) \vert k; m \rangle$.

Here, we want to make a stop and remember the action of a lossless $\mathrm{SU}(2)$ beam splitter \cite{Campos1989,Luis1995}, 
\begin{eqnarray}
\hat{T}(\theta)= \mathrm{exp} \left[ i \theta \left( \hat{a}_{1}^{\dagger}  \hat{a}_{2} + \hat{a}_{1}  \hat{a}_{2}^{\dagger}  \right) /2 \right],
\end{eqnarray}
such that for a 50/50 beam splitter, $\theta = \pi/2$, we can rewrite the $\mathrm{SU}(1,1)$ bases above, 
\begin{eqnarray}
\vert 1/2; m \rangle  &=&  \frac{(-i)^{m}}{m!} \hat{T}(\pi/2) \left( \hat{a}_{1}^{\dagger m} \hat{a}_{2}^{\dagger m} \right) \hat{T}^{\dagger}(\pi/2) ~\vert 0, 0 \rangle,  \nonumber \\
\vert 1; m \rangle_{01} &=& \frac{(-i)^{m}}{\sqrt{2 m!(m+1)!}} \hat{T}(\pi/2) \left( \hat{a}_{1}^{\dagger m+1} \hat{a}_{2}^{\dagger m} - i \hat{a}_{1}^{\dagger m} \hat{a}_{2}^{\dagger m+1} \right) \hat{T}^{\dagger}(\pi/2) ~\vert 0, 0 \rangle, \nonumber \\
\vert 1; m \rangle_{10} &=& \frac{(-i)^{m}}{\sqrt{2 m!(m+1)!}} \hat{T}(\pi/2) \left( \hat{a}_{1}^{\dagger m} \hat{a}_{2}^{\dagger m+1} - i \hat{a}_{1}^{\dagger m+1} \hat{a}_{2}^{\dagger m} \right) \hat{T}^{\dagger}(\pi/2) ~\vert 0, 0 \rangle,  \nonumber \\
\vert 3/2; m\rangle &=& \frac{(-i)^{m+1}}{\sqrt{2 m!(m+2)!}} \hat{T}(\pi/2) \left( \hat{a}_{1}^{\dagger m+2} \hat{a}_{2}^{\dagger m} - i \hat{a}_{1}^{\dagger m} \hat{a}_{2}^{\dagger m+2} \right) \hat{T}^{\dagger}(\pi/2) ~\vert 0, 0 \rangle, 
\end{eqnarray}
in terms of ideal 50/50 beam splitter states.
We have in our hands four phonon subspaces with an underlying $\mathrm{SU}(1,1)$ symmetry that resolves the Hilbert spaces of a lossless 50/50 $\mathrm{SU}(2)$ beam splitter.

Now, lets go back to the ideal evolution of the model and consider an initial state composed by the $m$-th state in any of the $\mathrm{SU}(1,1)$ subspaces and the ion in the excited state, $\vert \psi(0) \rangle = \vert k, m \rangle \vert e \rangle$. 
It is straightforward to see, 
\begin{eqnarray}
\vert \psi(t) \rangle = \cos g t \sqrt{(m+1)(m+2k)} ~\vert k; m \rangle \vert e \rangle - i \sin g t \sqrt{(m+1)(m+2k)}  ~\vert k; m + 1 \rangle \vert g \rangle,
\end{eqnarray}
that we can scale the $\mathrm{SU}(1,1)$ state ladder by a sequence of red second sideband driving and $\pi$-pulses, $R_{1}(\pi) = \mathrm{exp}\left( i \pi \sigma_{1} /2 \right)$, to switch the internal state of the ion,
\begin{eqnarray}
\begin{array}{lll}
t=0 				&:	& \vert k; 0 \rangle \vert e \rangle, \\
U_{2}\left(  \frac{\pi}{2 g\sqrt{2k} }  \right) 			&:	& \vert k; 1 \rangle \vert g \rangle, \\
e^{i \frac{\pi}{2} \sigma_{1} }	&:	& \vert k; 1 \rangle \vert e \rangle, \\
U_{2}\left(  \frac{\pi}{2g\sqrt{2(2k+1)}}  \right) 	 			&:	& \vert k; 2 \rangle \vert g \rangle, \\
e^{i \frac{\pi}{2} \sigma_{1} }	&:	& \vert k; 2 \rangle \vert e \rangle, \\
U_{2}\left(  \frac{\pi}{2g\sqrt{3(2k+2)}}  \right) 	 			&:	& \vert k; 3 \rangle \vert g \rangle, 
\end{array}  \label{eq:Algorithm}
\end{eqnarray}
and so on.
Now, this procedure generates entangled orthogonal vibrational states that are factorized from the internal structure of the ion. 
Again, this result suggests the use of this system as an interferometer that might provide information about the characteristics of an experiment that might be, in principle, different from those available through the superposition of squeezed states proposed before.

\begin{figure}[htbp]
	\centering
	\includegraphics[scale=1]{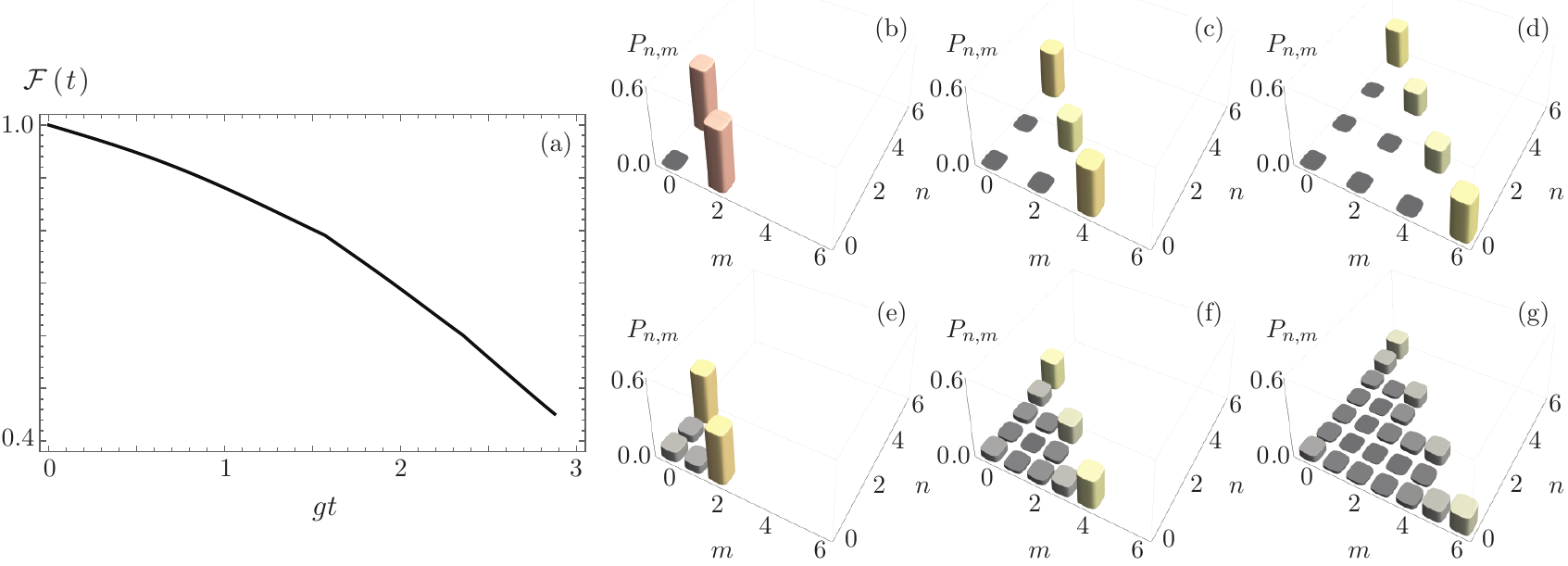}
	\caption{(a) Fidelity between ideal evolution and evolution under a common decay rate $\gamma = 0.1 ~g$ for the symmetric coupling case, $g_{1}= g_{2} = g$. The joint phonon number probability distribution for times where 50/50 $\mathrm{SU}(2)$ beam splitter states are expected under (b)-(d) ideal and (e)-(g) lossy evolution for the $\mathrm{SU}(1,1)$ subspace characterized by Bargmann parameter $k=1/2$ with (b),(e) $m=1$, (c),(f) $m=2$, and (d),(g) $m=3$.}
	\label{fig:Fig4}
\end{figure}

Figure \ref{fig:Fig4}(a) shows the fidelity, $\mathcal{F}(t)$, between the ideal evolution of the driving algorithm in Eq.(\ref{eq:Algorithm}) and lossy evolution under homogeneous decay for all components.
The initial state of the algorithm is the vacuum state for the subspace defined by the Bargmann parameter $k=1/2$.
That is, the initial state is the ion in the excited state and both vibrational modes cooled down to the vacuum state.
Figures \ref{fig:Fig4}(b) to \ref{fig:Fig4}(d) show the joint phonon number probability, $P_{n,m}$, at times where the 50/50 $\mathrm{SU}(2)$ beam splitter states are expected under ideal time evolution.
Figures \ref{fig:Fig4}(e) to \ref{fig:Fig4}(g) show the same probabilities for lossy evolution.

\section{Conclusion}

We proposed a trapped ion model under second sideband resolved blue and red driving of two orthogonal modes of the center mass motion in the Lamb-Dicke regime.
For parameter regimes providing an effective model with asymmetric coupling of the vibration modes with the internal state of the ion, we showed that time evolution of arbitrary vibrational states with balanced superposition of the internal states of the ion generates a so-called two-mode $\mathrm{SU}(1,1)$ Perelomov coherent state where the vibrational modes are separable.
Uneven superposition of the internal states of the ion produces a superposition where entanglement of the whole state of the ion arises.
We also showed that under red driving only, the effective model is able to generate 50/50 $\mathrm{SU}(2)$ beam splitter vibrational states factorized from the internal state of the ion.
The fact that it is ideally possible to engineer these states with reversible dynamics suggests the use of these models as interferometers to characterize real-world experiments.

\begin{acknowledgments}
C.H.A. acknowledges financial support from CONACYT doctoral grant No. 455378 and B.M.R.-L. from CONACYT CB-2015-01/255230 and CONACYT FORDECYT-296355. 
\end{acknowledgments}

%%%%%%%%%%%%%%%%%%%%%%% References %%%%%%%%%%%%%%%%%%%%%%%%%
%\bibliography{referencesBlas}


%
	
\end{document}